\documentclass[%
reprint,aps,prc,
superscriptaddress,
showpacs,preprintnumbers,
amsmath,
amssymb,
twolcolumn,
floatfix,  
]{revtex4-2}

\usepackage{lineno}  
\usepackage{lipsum}  

\usepackage{graphicx} 
\usepackage{dcolumn}
\usepackage{bm}
\usepackage{lmodern} 
\usepackage{ctable}
\usepackage{colortbl}
\usepackage{amsfonts}
\usepackage{amssymb}
\usepackage{gensymb}
\usepackage{caption}
\usepackage{sidecap}
\usepackage{hyperref}
\usepackage{xcolor}
\usepackage{float}
\usepackage{multirow}
\usepackage{caption}
\usepackage{subcaption}
\usepackage{tabularray}  
\usepackage{array}       
\usepackage{enumitem} 
\usepackage[normalem]{ulem} 
\usepackage{siunitx}
\usepackage{comment}

\usepackage[font=footnotesize]{caption}

\newcommand{\softsh}[1]{\texttt{#1}}
\newcommand{\soft}[1]{\texttt{#1}\,}

\newcommand{\lisepp}{\soft{LISE$^{++}$}}
\newcommand{\liseppsh}{\softsh{LISE$^{++}$}}

\newcommand{\liseppcutesh}{\softsh{LISE$^{++}_{cute}$}}

\newcommand{\nuc}[2]{\hbox{$^{#1}$#2}}

\begin{document}

\preprint{APS. Version 2.3}




\title{Quantifying uncertainty in physics-based predictions of rare-isotope production cross sections via Bayesian-inspired model averaging across nuclear mass tables}

\newcommand{\aNSCL}{\affiliation{National Superconducting Cyclotron Laboratory, Michigan State University, East Lansing, MI 48824, USA }}
\newcommand{\aMSUphys}{\affiliation{Department of Physics and Astronomy, Michigan State University, East Lansing, MI 48824, USA}}
\newcommand{\aMSUchem}{\affiliation{Department of Chemistry, Michigan State University, East Lansing, MI 48824, USA}}
\newcommand{\aFRIB}{\affiliation{Facility for Rare Isotope Beams, Michigan State University, East Lansing, MI 48824, USA}}

\newcommand{\aUMass}{\affiliation{University of Massachusetts Lowell, Lowell MA 01854}}
\newcommand{\aKorea}{\affiliation{Center for Exotic Nuclear Studies, Institute for Basic Science, Daejeon 34126, Republic of Korea}}
\newcommand{\aRIKEN}{\affiliation{RIKEN Nishina Center for Accelerator-Based Science, RIKEN, 2-1 Hirosawa, Wako, Saitama 351-0198, Japan}}

\author{O.~B.~Tarasov} \email{tarasov@frib.msu.edu}  \aFRIB




\date{\today}

\begin{abstract}
Accurate prediction of fragmentation cross sections is essential for rare-isotope beam production, planning new-isotope searches, and designing experiments to study the most exotic regions of the nuclear chart. However, existing reaction models and phenomenological cross-section parameterizations often exhibit significant deviations over broad regions of mass and charge. In this work, a Bayesian-inspired model-averaging framework is developed to combine abrasion--ablation (AA) calculations based on multiple nuclear mass tables into a single statistically weighted estimate. For the calibrated systems, the model weights are assigned empirically according to the relative quality of fit to measured cross sections, thereby reducing systematic model bias while preserving the underlying physics content of the AA description.

The weights are constrained using proton-rich fragmentation data for the $^{78}$Kr and $^{124}$Xe projectiles. The resulting parameter trends are then propagated to the $^{92}$Mo and $^{144}$Sm systems through a controlled scaling procedure. In the present implementation, the excitation-energy prescription is fixed, while the averaging is performed across nuclear-mass inputs; the framework provides both weighted cross sections and associated uncertainty estimates.

Applied to proton-rich fragmentation, the present approach provides a practical basis for interpolation and limited extrapolation in regions relevant to rare-isotope production. The resulting predictions are used to assess the production of very proton-rich nuclei, and candidate new isotopes are discussed.
\end{abstract}

\maketitle



\section{Introduction}

Although the nuclear landscape is expected to include roughly 8000 particle-bound nuclei, only about 3000 have been identified experimentally to date \cite{Neufcourt2020}. At intermediate and high energies, projectile fragmentation and fission of primary beams on production targets remain the dominant route for extending the chart of nuclides below uranium, also complemented by fusion–evaporation, including recent proton-rich discoveries in the mid-mass region $30\le Z \le 70$ \cite{59Ge_Ciemny,63Se_Blank,72Rb_Suzuki,81Mo_Suzuki,96In_Celikovich,124Xe_Suzuki25,Suzuki25_Z60,Suzuki25_Z63}.
Accurate prediction of projectile--fragmentation cross sections is therefore essential for rare-isotope beam production, separator optimization, and planning of experiments aimed at the discovery of new isotopes, especially near the proton and neutron drip lines where yields rapidly decrease and searches may rely on only a few observed events.

For fast-beam fragmentation, rare-isotope production cross sections are commonly estimated using the following approaches:
\begin{itemize}[itemsep=4pt, parsep=0pt, topsep=4pt, leftmargin=14pt]
    \item \emph{Phenomenological parameterizations} (e.g.~\cite{EPAX2,EPAX3,FRACS}):  
    Data-driven fits are convenient for first estimates but have limited predictive power when extrapolated into regions with little or no data and do not incorporate nuclear masses. They therefore primarily serve as a reference.

    \item \emph{Mass-based systematics} ($\Delta BE$~\cite{82Se_2025} and $Q_g$~\cite{76Ge_PRC}):  
    Simple mass--yield correlations capture broad trends and can highlight structural effects (e.g., odd--even staggering and local ``kinks''), but they are primarily constrained and validated on neutron-rich data sets, limiting their reliability outside that domain.
    
    \item \emph{Abrasion--ablation models}: 
    Physics-based two-step descriptions are more microscopic, but remain sensitive to model assumptions about excitation-energy generation and de-excitation dynamics and often require parameter optimization to reproduce measured cross sections \cite{82Se_2013,Kubeila21,78Kr_Suzuki25,82Se_2025}.
    
    \item \emph{Intranuclear-cascade (INC) plus statistical de-excitation}:  
    INC-based descriptions (e.g., INCL coupled to a statistical de-excitation model) are well established for nucleon- and light-ion--induced reactions \cite{INCL_Boudard2013,INCL_Cugnon2007,INCL-Liege}, whereas for heavy-ion fragmentation on production targets, the benchmarking is less systematic and results are more sensitive to the cascade--to--de-excitation interface, especially at intermediate energies.
    
\end{itemize}

The present work focuses on inclusive production cross sections for planning and optimization, for which the abrasion--ablation (AA) framework remains the standard physics-based baseline and has been extensively benchmarked against projectile-fragmentation systematics. Consequently, the AA framework is adopted as the reference, and residual uncertainties associated primarily with the choice of nuclear masses are treated statistically using a Bayesian-inspired model-averaging procedure with empirically assigned model weights.

Recent high-quality measurements provide stringent constraints on fragmentation models. In particular, cross sections for \nuc{78}{Kr}+\nuc{9}{Be}~\cite{78Kr_Suzuki25} and \nuc{124}{Xe}+\nuc{9}{Be}~\cite{124Xe_Suzuki25} probe proton-rich nuclei directly relevant to new-isotope discovery programs and reveal systematic discrepancies with existing calculations. These data sets span broad ranges in $Z$ and $A$ and serve as the primary experimental inputs in the present analysis.

The \liseppcutesh~\cite{LISE2016,LISE2023} implementation of the abrasion--ablation (AA) model~\cite{LISEAA} has been applied previously to \nuc{78}{Kr}+\nuc{9}{Be} \cite{Kubeila21,78Kr_Suzuki25}. In those studies, a Gaussian-shaped prefragment excitation-energy distribution improved the reproduction of measured isotopic yields compared to simpler prescriptions. However, the calculations were restricted to a particular AA realization---most notably, one selected mass table among several available options---and residual discrepancies remained, underscoring that no single AA realization is globally reliable across all isotopic regions.

Motivated by these observations, a Bayesian-inspired model-averaging framework is employed in conjunction with abrasion--ablation calculations performed with \liseppsh. Predictions from multiple AA realizations based on different nuclear mass tables are combined, and their statistical weights are determined using the $^{78}$Kr+$^{9}$Be and $^{124}$Xe+$^{9}$Be data. Alternative excitation-energy prescriptions were benchmarked within \liseppsh, but the present analysis adopts a single excitation-energy parameterization, kept fixed throughout all minimizations and propagated calculations. The resulting model-averaged (MA) cross sections and quantified uncertainties provide a practical basis for interpolation and limited extrapolation toward unmeasured nuclei and for planning future experiments, including new-isotope searches. The framework is applied to proton-rich isotope production induced by \nuc{78}{Kr}, \nuc{92}{Mo}, \nuc{124}{Xe}, and \nuc{144}{Sm}, with weights constrained by the \nuc{78}{Kr} and \nuc{124}{Xe} measurements and extended to \nuc{92}{Mo} and \nuc{144}{Sm} via controlled scaling of the fitted parameters.


\section{Abrasion--ablation framework and fitting methodology}

Projectile fragmentation at intermediate and high energies is commonly described in the abrasion--ablation (AA) framework \cite{Gaimard91}. The fast abrasion step determines the excited prefragment $(A^{*},Z^{*})$ and its excitation energy, while the subsequent ablation step yields the final fragment $(A,Z)$ through statistical de-excitation via particle evaporation and, when relevant, fission.
The \lisepp\ implementation employs a geometrical abrasion picture and an analytical treatment that enables rapid evaluation of inclusive production cross sections over broad regions of $Z$ and $A$ (see Ref.~\cite{Kubeila21} for implementation details). This computational efficiency is particularly important for rare-isotope applications, where cross-section estimates are required for nuclei far from stability.

The \lisepp\ minimization utility uses measured cross sections to constrain selected AA parameters by minimizing an objective function defined from experimental residuals. Relative to purely phenomenological parameterizations, this approach provides a physics-based description that can be systematically improved by varying nuclear inputs and model prescriptions. At the same time, residual discrepancies and model dependence remain, driven in particular by assumptions on excitation-energy generation and by the choice of nuclear masses. These sensitivities motivate both robust fitting strategies and the statistical combination of multiple AA realizations used in the MA analysis.

Because the excitation-energy prescription is an important source of model dependence, several functional forms were examined within \liseppsh. The implemented excitation-energy options and the rationale for the prescription adopted in the present work are summarized below. In the present analysis, one excitation-energy parameterization is selected and then used consistently for all minimizations and propagated calculations.

\subsection{Prefragment excitation-energy models}

In earlier AA-based studies of the $^{78}$Kr fragmentation data~\cite{Kubeila21,78Kr_Suzuki25}, 
the prefragment excitation-energy distribution was parameterized by a Gaussian.
This choice was motivated by the statistical hole-energy (diabatic) picture of Ref.~\cite{Gaimard91}, where the abrasion process is assumed to be fast enough that the single-particle orbits of the spectator are essentially unchanged. 
In that framework, removing a single nucleon creates a vacancy (a ``hole'') below the Fermi surface; for a Woods--Saxon potential the hole energy spans roughly $0$--$40$~MeV (for $E_F\approx -7.4$~MeV). 
Assuming an equal probability to generate a hole in each single-particle level and using a level density $g(E)\propto E$ 
leads to a one-hole excitation-energy distribution that is approximately linear in $E^{*}$ (a ``triangular'' form). 
For larger mass losses, the total excitation energy is obtained as the convolution of several such single-hole contributions, 
which produces broader, increasingly Gaussian-like distributions (Fig.~2 of Ref.~\cite{Gaimard91}), 
thus motivating the Gaussian approximation used in practice.

To test alternative behaviors suggested by other approaches, \lisepp was extended to include additional functional forms for the prefragment excitation-energy distribution, including the INC exponential-deposition model (EDM) used in Ref.~\cite{Audirac13} and a log-normal form motivated by intranuclear-cascade event generators (e.g.\ \soft{BeAGLE}~\cite{Ferrari1996}, \soft{INCL-Li\`ege}~\cite{INCL-Liege}).
In addition, an ``exponential-hole'' variant of the Gaimard--Schmidt convolution scheme was implemented in \liseppsh, in which the one-hole excitation-energy spectrum $P_1(E^{*})$ is modeled as an exponentially damped triangle,
\begin{equation}
P_1(E^{*}) \propto E^{*}\,\exp(-E^{*}/E_0)\qquad (E^{*}<E_h),
\end{equation}
with $E_h$ the hole-depth cutoff and $E_0$ a damping (e-folding) energy scale that controls the suppression of deep holes. The parameter $E_0$ is treated as tunable. The exponential factor reduces the high-$E^{*}$ tail relative to a purely Gaussian parameterization. Notably, convolving these damped one-hole distributions yields multi-hole excitation-energy distributions that are very close in shape to both the exponential-deposition and log-normal parameterizations.

Beyond its improved benchmarking performance relative to a Gaussian excitation-energy shape for several primary beams (mainly observed for neutron-rich cases such as Refs.~\cite{198Pt_NSCL,82Se_2025}), the EDM  also offers a more practical and physically transparent scaling with projectile mass. 
In particular, the excitation-energy scale is governed primarily by a single parameter (the characteristic scale $T$), whereas a Gaussian description introduces an additional width parameter (a second-order moment) whose scaling is less constrained and therefore difficult to extrapolate reliably between projectiles.

In \liseppsh, the mean excitation-energy scale $T$ used in the EDM~\cite{Audirac13} is parameterized as
\begin{equation}
T = k_1 + k_2 \Delta A_{\text{abr}} + k_{NZ} \left(\Delta N_{\text{abr}} - \Delta Z_{\text{abr}}\right),
\label{eq:Tmean}
\end{equation}
where $\Delta A_{\text{abr}}$, $\Delta N_{\text{abr}}$, and $\Delta Z_{\text{abr}}$ are the numbers of abraded nucleons, neutrons, and protons from the projectile, respectively.


\subsection{AA model settings and fit parameters}
\label{sec:AA-settings}

Calculations were performed with \liseppcutesh~v17.18. The AA settings used in the minimization were:
\begin{itemize}[itemsep=2pt, parsep=0pt, topsep=2pt, leftmargin=14pt]
\item Evaporation-distribution dimension: 64
\item Decay modes: $n$, $p$, $\alpha$, break-up
\item Daughter excitation-energy mode: fast
\item Excitation-energy model: \soft{EDM} (\lisepp model \#4)
\end{itemize}
In the abrasion stage, \lisepp\ employs the standard geometrical participant--spectator (sharp-cutoff) picture of abrasion--ablation fragmentation.

The parameters varied during the minimization were:
\begin{itemize}[itemsep=2pt, parsep=0pt, topsep=2pt, leftmargin=14pt]
\item EDM parameters $k_1$, $k_2$, and $k_{NZ}$ (Eq.~\ref{eq:Tmean})
\item Evaporation odd--even parameter $\delta_{oe}$
\item Global scaling factor $S_{\sigma}$
\item Effective Coulomb-barrier parameter $\Delta R$
\end{itemize}
Further definitions and defaults are given in Appendix~\ref{app:AA_settings}.

\subsection{Objective function for AA minimization}
\label{sec:objective}

For each nuclear mass model, the AA parameters were obtained by minimizing an objective function
$\mathcal{J}$ constructed from the residuals between calculated and measured cross sections.
The objective function combines two complementary contributions:
(i) a \emph{quadratic} metric, $\mathcal{L}_{\mathrm{quad}}$, and
(ii) a \emph{logarithmic} metric, $\mathcal{L}_{\log}$.
Together, these terms balance sensitivity to well-bound nuclei and to the low-cross-section tails
associated with the most exotic isotopes. Specifically,
\begin{equation}
\mathcal{J}=a\,\mathcal{L}_{\mathrm{quad}}+(1-a)\,\mathcal{L}_{\log},
\label{eq:J_def}
\end{equation}
where the mixing parameter $a$ was chosen empirically so that, in practice, both components contribute
comparably (in absolute magnitude) to the minimization.

\paragraph{Quadratic component.}
The quadratic metric emphasizes agreement in absolute cross-section values and is therefore more sensitive
to stable or near-stable isotopes:
\begin{equation}
\mathcal{L}_{\mathrm{quad}} = \frac{1}{N}\sum_{i=1}^{N}\left(\frac{\sigma_{\mathrm{exp},i}-\sigma_{\mathrm{calc},i}}{\sigma_{\mathrm{err},i}}\right)^2 .
\label{eq:Lquad}
\end{equation}
Here $\sigma_{\mathrm{err},i}$ is the experimental uncertainty of the measured cross section $\sigma_{\mathrm{exp},i}$.

\paragraph{Logarithmic component.}
The logarithmic metric quantifies discrepancies on a multiplicative scale and is less dominated by the highest
cross sections, making it comparatively more sensitive to the low-cross-section tails. A commonly used reference
form is the mean absolute logarithmic residual,
\begin{equation}
\mathcal{L}_{\log}^{(|\log|)} = \frac{1}{N}\sum_{i=1}^{N}\left|\log_{10}\!\left(\frac{\sigma_{\mathrm{calc},i}}{\sigma_{\mathrm{exp},i}}\right)\right| .
\label{eq:Llog_abs}
\end{equation}

In the \lisepp\ AA minimization, a weighted, soft-capped logarithmic deviation term is employed to (i) include cross-section
uncertainties when available and (ii) limit the leverage of extreme outliers. Defining
\begin{equation}
d_i=\log_{10}\!\left(\frac{\sigma_{\mathrm{calc},i}}{\sigma_{\mathrm{exp},i}}\right),\qquad
w_i=\frac{\sigma_{\mathrm{exp},i}}{w_{1,i}},
\label{eq:di_wi_def}
\end{equation}
the logarithmic metric is evaluated as
\begin{equation}
\mathcal{L}_{\log} = \frac{\sum_i w_i\,S(|d_i|)}{\sum_i w_i},
\label{eq:Llog_soft}
\end{equation}
where $S$ behaves like $10^{|d|}$ for typical deviations but is softly capped for large $|d|$ to reduce the leverage
of outliers (Appendix~\ref{app:softpow10}).
Here $w_i$ denotes a \emph{data-point} weight used in the objective function and should not be confused with the \emph{model-averaging} weights $w_m$ introduced later. \vspace{0.2 cm}

\paragraph{Effective uncertainty for logarithmic weighting.}
The effective uncertainty $w_{1,i}$ is defined as
\begin{equation}
w_{1,i}=\max\!\left(\tilde w_{1,i},\,\sigma_{\min}\right),
\label{eq:w1_def}
\end{equation}
with $\tilde w_{1,i}=\Delta\sigma_i$ when experimental uncertainties are available and
$\tilde w_{1,i}=f_{\rm err}\,\sigma_{\mathrm{exp},i}$ otherwise; $\sigma_{\min}$ is a fixed floor that prevents
pathologically large weights. For constant fractional uncertainty (above the floor, $w_{1,i}\propto\sigma_{\mathrm{exp},i}$),
$w_i=\sigma_{\mathrm{exp},i}/w_{1,i}\approx\mathrm{const}$ and Eq.~\eqref{eq:Llog_soft} assigns comparable
influence across the dynamic range, including the low-cross-section tails. When the uncertainties deviate from a constant fractional form, the weights vary accordingly and reflect the reported point-to-point uncertainties.


\subsection{Robust AA minimization for BMA-inspired fits}

\paragraph{Local-line cycling (mitigating local minima).}
In the present AA minimization, the objective function is written as a weighted sum of four terms: global $\chi^{2}$, global log$_{10}$-difference (LoD), local $\chi^{2}$, and local LoD, where ``local'' refers to a user-selected isotopic line (e.g., fixed $Z$ or fixed $N$). If the local terms are given large weights, the minimizer can overfit the selected line while leaving neighboring lines poorly reproduced. Nonlinear least-squares minimization was performed with the Levenberg--Marquardt algorithm using the \texttt{levmar} C/C++ implementation \cite{lourakis04LM}.

To reduce this tendency, the implementation provides an option to \emph{cycle} the local line within a single minimization run. During cycling, the local subset is periodically shifted to neighboring $Z$ (or $N$) values and then returned to the originally selected line. Fit parameters are updated continuously throughout the cycle and are never reset. Cycling is controlled by (i) the scan range $\Delta$ (no cycling for $\Delta=0$; otherwise include neighbors up to $\pm\Delta$) and (ii) the number of objective-function evaluations performed at each line position. While cycling does not guarantee convergence to the global minimum, it reduces locking to minima centered on a single $Z$/$N$ row and improves the smoothness and robustness of convergence in difficult fits.

\paragraph{Diagonal scaling (\soft{DSCL}) in \soft{levmar}.}
AA fits involve parameters with very different numerical scales (e.g., energies/temperatures in MeV, factors of order unity, and small correction terms), which can lead to poor conditioning of $J^{T}J$, unbalanced step sizes, and occasional near-singular behavior in the minimization. To mitigate this, \lisepp\ offers an optional diagonal scaling (\soft{DSCL}) that is passed to the \soft{levmar} engine as a per-parameter scaling vector.

For each bounded parameter $p_i\in[p_i^{\min},p_i^{\max}]$, the span is computed as $\Delta p_i = |p_i^{\max}-p_i^{\min}|$ and the scaling factor is set to $d_i = 1/\Delta p_i$, with a safe fallback to $d_i=1$ if the span is zero or invalid. LM then operates on scaled variables $q_i = d_i\,p_i$, bringing bounded parameters to comparable, order-of-unity scales internally. In practice, \soft{DSCL} reduces anisotropy in parameter space, makes $J^{T}J$ less prone to near-singularity, and improves stability and convergence smoothness. As a numerical preconditioner, \soft{DSCL} does not alter the objective function; its impact may be limited when convergence is dominated by strong parameter correlations or poorly chosen bounds, and extreme scaling (very small $\Delta p_i$) can lead to conservative steps. When \soft{DSCL} is disabled, the historical behavior is restored.

\section{\label{sec:78Kr_data} Constraining the AA model with the \nuc{78}{Kr} and \nuc{124}{Xe} data}

The minimization was first carried out for the \nuc{78}{Kr} data of Ref.~\cite{78Kr_Suzuki25} in the charge range $26 \le Z \le 36$.
For each of the 12 mass models considered here, the fit parameters defined in Sec.~\ref{sec:AA-settings}, including the excitation-energy parametrization of Eq.~\ref{eq:Tmean}, were varied in the minimization.
The resulting minimum values of the objective function $\mathcal{J}$ and the corresponding best-fit parameter sets are summarized in Table~\ref{tab:78kr}.

\begin{table}[h]
\setlength{\abovecaptionskip}{10pt}
\centering
\setlength{\extrarowheight}{1.5pt}
\setlength{\tabcolsep}{2.0pt}
\caption{Results of the Abrasion--Ablation minimization for \nuc{78}{Kr} projectile-fragmentation products. For each mass model, the table lists the minimum objective-function value $\mathcal J$ and the corresponding best-fit parameters. The last two rows summarize the weighted-average values and their uncertainties.}
\label{tab:78kr}
\small
\begin{ruledtabular}
\begin{tabular}{lc|c|
>{\footnotesize}r
>{\footnotesize}r
>{\footnotesize}r|
>{\footnotesize}r
>{\footnotesize}r
>{\footnotesize}r}
\multicolumn{2}{c|}{\text{Mass model}} & $\mathcal J$ & $k_1$ & $k_2$ & $k_{NZ}$ & $S_{\sigma}$ & $\delta_{oe}$ & $\Delta R$ \\
\hline
AME2020   & \cite{AME2020}   & 1.40 & 12.3 & -0.39 &  0.27 & 1.1 &  8.6 & 5.9 \\
FRDM2012  & \cite{PM-PRL12}  & 1.63 & 11.9 & -0.48 &  0.50 & 1.5 &  7.7 & 6.5 \\
HFB-22    & \cite{SG-PRC13}  & 3.01 & 14.7 & -0.26 &  0.16 & 1.1 &  9.0 & 6.0 \\
HFB-27    & \cite{SG-PRC13}  & 2.06 & 14.5 & -0.50 &  0.40 & 0.8 &  6.9 & 6.3 \\
KTUY      & \cite{KTUY-PTP05}& \textbf{1.10} & 12.7 & -0.46 &  0.39 & 1.3 &  8.8 & 6.5 \\
LISE-LDM3 & \cite{LISE08}    & 2.64 & 14.1 & -0.33 &  0.38 & 1.4 &  8.0 & 6.0 \\
NL3*      & \cite{rN3}       & 3.78 & 15.1 &  0.20 & -0.18 & 1.1 & 10.0 & 6.0 \\
SLy4      & \cite{SLy4}      & 2.30 & 16.3 & -0.12 &  0.32 & 1.2 & 10.9 & 5.9 \\
SV-min    & \cite{SVmin}     & 2.27 & 13.8 &  -0.03 &  0.40 & 1.0 & 11.6 & 6.0 \\
TUYY      & \cite{TTYY}      & 2.19 & 12.9 & -0.40 &  0.30 & 1.0 &  7.2 & 6.0 \\
UNEDF1    & \cite{UNEDF1}    & 3.19 & 11.5 &  0.06 &  0.39 & 0.5 & 10.6 & 6.1 \\
WS4$_{\mathrm{RBF}}$ & \cite{WS4RBF} & 1.80 & 13.0 & -0.49 &  0.41 & 0.8 &  7.2 & 6.0 \\
\hline
\multicolumn{3}{r|}{\textit{Weighted average \hspace*{0.4cm}}}  & 13.1 & -0.37 & 0.36 & 1.1 & 8.6 & 6.2 \\
\multicolumn{3}{r|}{\textit{Uncertainty \hspace*{0.4cm}}}    &  1.1 &  0.16 & 0.11 & 0.2 & 1.2 & 0.2 \\
\end{tabular}
\end{ruledtabular}
\end{table}

\begin{table}[h]
\setlength{\abovecaptionskip}{10pt} 
\centering
\setlength{\extrarowheight}{1.5pt}
\setlength{\tabcolsep}{2.0pt} 

\centering
\caption{Same as Table~\ref{tab:78kr}, but for the \nuc{124}{Xe}+Be fragmentation data and the corresponding AA minimization results.}
\label{tab:124xe}
\small
\begin{ruledtabular}
\begin{tabular}{lc|c|
>{\footnotesize}r
>{\footnotesize}r
>{\footnotesize}r|
>{\footnotesize}r
>{\footnotesize}r
>{\footnotesize}r}
\multicolumn{2}{c|}{\text{Mass model}} & $\mathcal J$ & $k_1$ & $k_2$ & $k_{NZ}$ & $S_{\sigma}$ & $\delta_{oe}$ & $\Delta R$ \\
\hline
AME2020  &\cite{AME2020}& 2.80 & 23.9 & -0.50 &  0.50 & 0.5 & 12.0 & 4.2 \\
FRDM2012 &\cite{PM-PRL12}& 3.61 & 21.5 & -0.36 &  0.30 & 0.5 &  9.4 & 4.3 \\
HFB-22    &\cite{SG-PRC13}& 3.00 & 23.1 & -0.13 & -0.27 & 1.8 & 12.2 & 5.6 \\
HFB-27    &\cite{SG-PRC13}& \textbf{1.49} & 21.8 & -0.20 & -0.09 & 0.8 & 11.7 & 4.7 \\
KTUY     & \cite{KTUY-PTP05}& 2.03 & 21.2 & -0.45 &  0.41 & 0.8 & 13.0 & 4.7 \\
LISE-LDM3 & \cite{LISE08}& 2.86 & 20.5 & -0.47 & -0.02 & 1.6 & 11.2 & 5.0 \\
NL3*     & \cite{rN3} & 8.46 & 19.6 & -0.11 &  0.30 & 1.8 & 12.5 & 5.7 \\
SLy4     & \cite{SLy4} & 5.08 & 21.9 & -0.44 &  0.50 & 1.6 & 13.0 & 5.3 \\
SV-min   & \cite{SVmin}& 3.28 & 25.0 & -0.29 &  0.50 & 0.8 & 12.9 & 4.3 \\
TUYY     &  \cite{TTYY}& 3.23 & 20.8 & -0.31 &  0.17 & 1.0 & 12.8 & 5.0 \\
UNEDF1   & \cite{UNEDF1}& 2.17 & 19.7 & -0.31 &  0.32 & 1.5 & 13.0 & 5.2 \\
WS4$_{\mathrm{RBF}}$ & \cite{WS4RBF} & 2.09 & 19.5 & -0.11 &  0.00 & 0.9 & 12.8 & 4.8 \\
\hline
\multicolumn{3}{r|}{\textit{Weighted average \hspace*{0.4cm}}}  & 21.4 & -0.3 & 0.1 & 1.0 & 12.3 & 4.8 \\
\multicolumn{3}{r|}{\textit{Uncertainty \hspace*{0.4cm}}}    &  1.4 &  0.1 & 0.2 & 0.4 & 0.8 & 0.3 \\
\end{tabular}
\end{ruledtabular}
\end{table}

The results listed in Table~\ref{tab:78kr} show a clear dependence on the adopted mass model.
Among the 12 cases considered, the KTUY masses provide the best overall description of the \nuc{78}{Kr} data, yielding the lowest objective-function value, $\mathcal{J}=1.10$, while AME2020 also gives a comparably good fit.
At the same time, the extracted fit parameters remain reasonably consistent across the full set of models.

\begin{figure*}[htbp]
\centering
\begin{subfigure}[t]{0.495\textwidth}
  \centering
  \includegraphics[width=\linewidth]{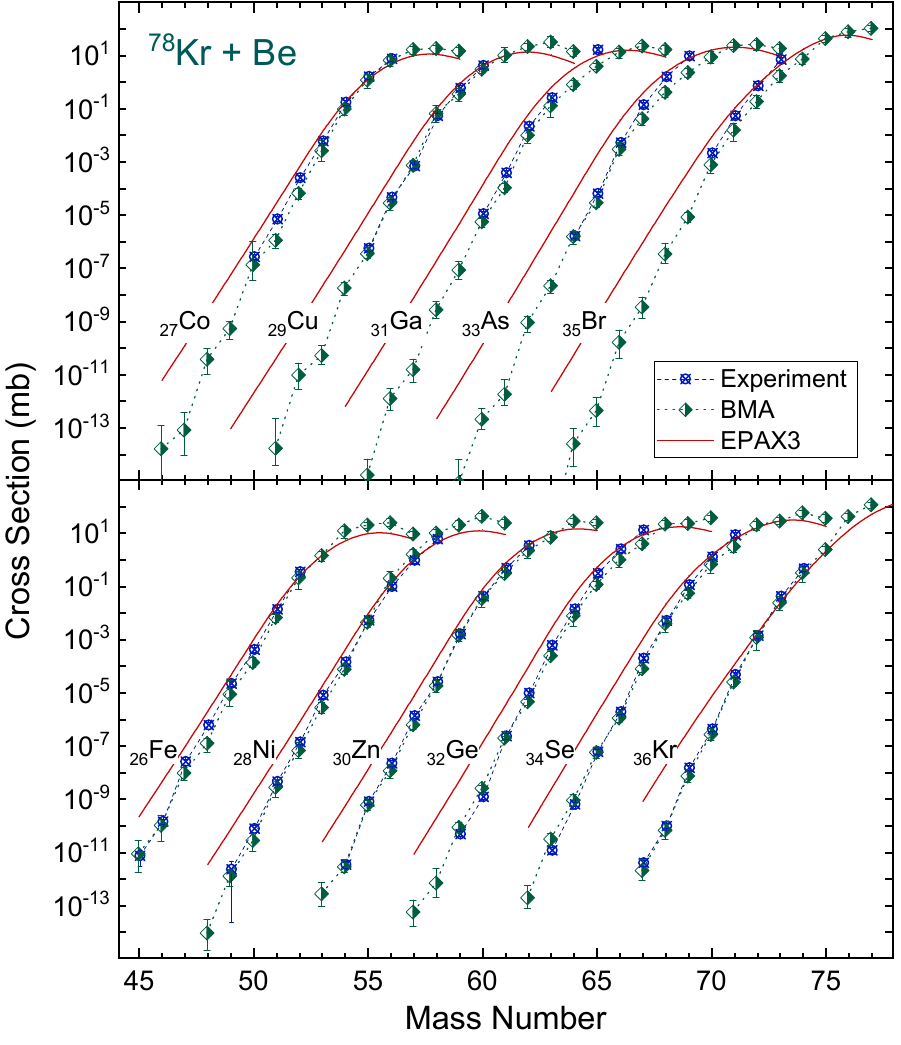}
\end{subfigure}
\hfill
\begin{subfigure}[t]{0.495\textwidth}
  \centering
  \includegraphics[width=\linewidth]{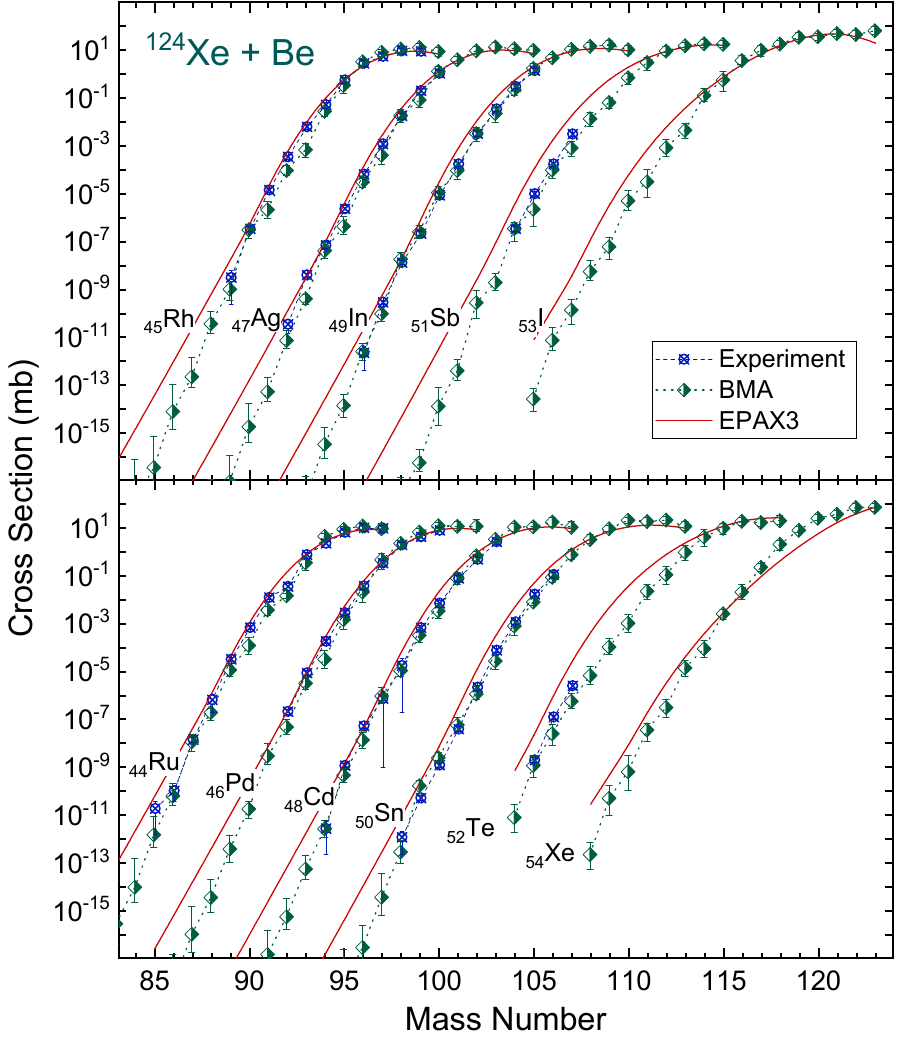}
\end{subfigure}
\caption{\label{fig:BMAcompare}
Measured production cross sections compared with the model-averaged abrasion--ablation (BMA-inspired) calculations and the EPAX3 parametrization. The isotopic distributions are shown as a function of mass number on a logarithmic cross-section scale. The left panel shows the \nuc{78}{Kr}+\nuc{9}{Be} reaction for fragments in the fitted range $26 \le Z \le 36$, while the right panel shows the \nuc{124}{Xe}+\nuc{9}{Be} reaction for fragments in the fitted range $44 \le Z \le 54$.}
\end{figure*}

The weighted-average value $k_1=13.1$ remains close to the values obtained previously with the Gaussian-shape excitation-energy model~\cite{Kubeila21,78Kr_Suzuki25}. Because $k_1$ is associated with the first-order excitation energy (in MeV) per abraded nucleon, this indicates that the leading term of the excitation-energy parametrization is locally well constrained. The parameters $k_2$ and $k_{NZ}$ also exhibit reasonably consistent trends across the tested mass models.
Their interpretation, however, remains primarily phenomenological within the present framework, and broader conclusions will require a more global analysis covering additional projectile systems and wider mass regions, since the present trends are inferred only from the restricted fragment range $26 \le Z \le 36$ on the proton-rich side.

In particular, $\Delta R$ shows only weak variation and remains close to the default value of 6~fm, which was explicitly extracted by Benlliure \textit{et al.} from fragmentation-corridor systematics~\cite{Benlliure1998}.
This indicates that the present data do not require any substantial modification of the geometrical radius term.

By contrast, the default choice $\delta_{oe}=12$~MeV in the AA model originates from the Ignatyuk pairing-gap parametrization~\cite{ignatyuk_istekov_smirenkin_1979}, $\Delta_0 \approx 12/\sqrt{A}$, adopted in Ref.~\cite{Gaimard91}.
Because this coefficient was introduced from level-density systematics for heavy nuclei, it should be regarded as a global default rather than a universal constant.
A similar trend toward a reduced odd--even correction was also observed in the recent analysis of the \nuc{82}{Se} data~\cite{82Se_2025}, where an even smaller value of about 6 was obtained.
Taken together, these observations support the use of the weighted-average values as a representative summary of the constraints derived from the \nuc{78}{Kr} data.

Finally, the fitted values of $S_{\sigma}$ remain close to unity, indicating that no substantial global renormalization of the calculated cross sections is required.
The weighted-average value $S_{\sigma}\approx 1.1$ therefore points to only a small residual normalization offset between the AA calculations and the measured yields.

The same minimization procedure was then applied to the \nuc{124}{Xe}+\nuc{9}{Be} data of Ref.~\cite{124Xe_Suzuki25} over the charge range $44 \le Z \le 54$, and the corresponding results are summarized in Table~\ref{tab:124xe}.
As for \nuc{78}{Kr}, the minimum value of $\mathcal{J}$ depends on the adopted mass model, with HFB-27 giving the lowest value among the models considered. The fitted parameters remain reasonably stable overall. In particular, $\delta_{oe}$ increases, as expected for the heavier \nuc{124}{Xe} system, and returns close to the default AA value of 12~MeV, whereas $\Delta R$ shows a slight decrease relative to the nominal 6~fm value. Although the weighted-average value of $S_{\sigma}$ remains close to unity, its spread across the mass models is larger than for the other fitted parameters. This suggests that, for the \nuc{124}{Xe} data, the main adjustment is an overall normalization of the calculated cross sections rather than a substantial change in the geometrical or odd--even terms.


\subsection{BMA of the \nuc{78}{Kr} and \nuc{124}{Xe} data}
\label{sec:BMA_78Kr_124Xe}
Bayesian-inspired model-averaged quantities were obtained using an empirical weighting scheme based on the relative quality of fit for each mass model. For each model $i$, the weight $w_i$ was taken to be inversely proportional to the objective-function value $\mathcal{J}_i$ and then normalized over all models,
\begin{equation}
w_i=\frac{J_i^{-1}}{\sum_k J_k^{-1}} .
\end{equation}
This prescription assigns greater weight to better-fitting models while smoothly reducing the contribution of less favorable ones. This empirical scheme differs from standard Bayesian model averaging in that the weights are not posterior model probabilities derived from an explicit likelihood-and-prior construction, but are instead assigned heuristically from the objective-function values.

For each nucleus, the cross sections predicted by the different mass-model calculations were combined using this weighting procedure. The averaging was performed in $\log_{10}$ space in order to accommodate the wide dynamic range of the calculated cross sections. Each model prediction was assigned a small symmetric relative uncertainty in linear space; after transformation to $\log_{10}$ space, this uncertainty becomes asymmetric. The combined central value was then obtained as a weighted mean of the log-transformed cross sections. The lower and upper uncertainties were evaluated separately, yielding an asymmetric uncertainty band in log space. The resulting central value and uncertainty bounds were finally transformed back to linear space, giving the MA cross section with asymmetric uncertainties.

The BMA-inspired cross sections obtained for the \nuc{78}{Kr}+\nuc{9}{Be} and \nuc{124}{Xe}+\nuc{9}{Be} systems are compared in Fig.~\ref{fig:BMAcompare} with the experimental data~\cite{78Kr_Suzuki25,124Xe_Suzuki25} and the EPAX3 parametrization~\cite{EPAX3}. The figure provides a compact overview of the isotopic distributions across the fragment chains included in the fits and highlights the overall agreement achieved by the BMA-inspired AA description.

The quantitative comparison between the BMA-inspired and measured cross sections for the \nuc{78}{Kr}+\nuc{9}{Be} and \nuc{124}{Xe}+\nuc{9}{Be} reactions is summarized in Table~\ref{tab:BMAcomp}. For \nuc{78}{Kr}, the MA values show a moderate overall overprediction, whereas for \nuc{124}{Xe} a moderate overall underprediction is observed. In both cases, the agreement is best for the central fitted fragment chains, while the largest deviations appear toward the edges of the fitted isotopic distributions.

\begin{table}[htbp]
\setlength{\abovecaptionskip}{8pt}
\setlength{\extrarowheight}{2.0pt}
\setlength{\tabcolsep}{2.0pt}
\centering
\caption{\label{tab:BMAcomp}
Summary of the agreement between the BMA-inspired and measured cross sections for the \nuc{78}{Kr}+\nuc{9}{Be} and \nuc{124}{Xe}+\nuc{9}{Be} reactions in the fitted charge intervals shown in Fig.~\ref{fig:BMAcompare}.}
\small
\begin{ruledtabular}
\begin{tabular}{lcc}
Quantity & \nuc{78}{Kr}+\nuc{9}{Be} & \nuc{124}{Xe}+\nuc{9}{Be} \\
\hline
Fitted $Z$ range & $26 \le Z \le 36$ & $44 \le Z \le 54$ \\
Within factor of 2 & 64\% & 56\% \\
Geom.-mean ratio $\sigma_{\mathrm{BMA}}/\sigma_{\mathrm{exp}}$ & 1.9 & 0.58 \\
Within $1\sigma$ & 61\% & 58\% \\
Within $2\sigma$ & 96\% & 76\% \\
\end{tabular}
\end{ruledtabular}
\end{table}

The small nonzero MA cross sections shown in Fig.~\ref{fig:BMAcompare} for slightly proton-unbound nuclei are a formal consequence of the AA model with tunneling enabled ($\Delta R \approx 6$ fm) and should be interpreted only as calculated populations of near-threshold resonant states, not as measurable transmitted fragment yields.


\section{Extension to the \nuc{92}{Mo} and \nuc{144}{Sm} systems \label{section_92Mo}}

\subsection{Propagation of the excitation-energy scale to additional projectiles}
\label{sec:scale_e0}

To propagate the excitation-energy scale from the calibrated projectiles to additional systems, a
\emph{two-point calibration of a separable size--isospin scaling law} is employed. Here
$\varepsilon(A,Z)$ denotes the projectile-dependent excitation-energy \emph{scale parameter} used in
the \soft{EDM} prescription (Eq.~\ref{eq:Tmean}). The projectile dependence is assumed to factorize
into a smooth, geometry-like size dependence and a small linear ``isospin tilt'' in the asymmetry
variable $I(A,Z)=(N-Z)/A$ (a first-order expansion in $I$):
\begin{equation}
\varepsilon(A,Z)=K\, f(A)\,\bigl[1 + C\, I(A,Z)\bigr].
\label{eq:eps_scaling}
\end{equation}
A radius-like baseline is adopted,
\begin{equation}
f(A)=A^{1/3}.
\label{eq:fA}
\end{equation}
This dependence follows the geometric scaling of the projectile radius $R\propto A^{1/3}$,
providing a natural, weakly varying size dependence for energy deposition in abrasion-based
pictures and avoiding overly strong $A$ trends in extrapolations.

Using two calibration projectiles with fitted excitation-energy scales
$y_1\equiv\varepsilon(A_1,Z_1)$ and $y_2\equiv\varepsilon(A_2,Z_2)$, the parameters $(K,C)$ are
determined for each mass model by solving the corresponding $2\times 2$ system
(Appendix~\ref{app:twopoint_scale}). The excitation-energy scale for an additional projectile
$(A_3,Z_3)$ is then obtained from Eq.~\eqref{eq:eps_scaling} evaluated for $(A_3,Z_3)$, with
$I_3=(N_3-Z_3)/A_3$.

In practice, this scaling is applied independently (for each mass model) to the \soft{EDM}
coefficients $k_1$, $k_2$, and $k_{NZ}$ in Eq.~\ref{eq:Tmean}, yielding propagated values for the
\nuc{92}{Mo} and \nuc{144}{Sm} projectiles based on the calibrated \nuc{78}{Kr} and \nuc{124}{Xe}
AA fits.


\subsection{Propagation of BMA weights to \nuc{92}{Mo} and \nuc{144}{Sm}}
\label{sec:weights_scale}

For each nuclear-mass model $m$, the AA minimizations constrained by the
\nuc{78}{Kr}+\nuc{9}{Be} and \nuc{124}{Xe}+\nuc{9}{Be} data yield objective-function values
$\mathcal{J}^{(78)}_{m}$ and $\mathcal{J}^{(124)}_{m}$ (Sec.~\ref{sec:78Kr_data}).
Because comparable calibration data are not available for \nuc{92}{Mo} and \nuc{144}{Sm},
projectile-dependent mass-model weights are inferred by interpolating the goodness-of-fit
between the two calibrated projectiles in the $(A,I)$ plane, where $I=(N-Z)/A$.

The scaled distance between a projectile $P$ and a calibrator $i\in\{78,124\}$ is defined as
\begin{equation}
D_{P,i}=
\sqrt{\left(\frac{|A_P-A_i|}{A_0}\right)^2+\lambda\left(\frac{|I_P-I_i|}{I_0}\right)^2},
\label{eq:dist_AI}
\end{equation}
where $A_0$ and $I_0$ set the relative scales of the mass and asymmetry differences and
$\lambda$ controls the weight of the asymmetry term. In the calculations reported here, $A_0=40$, $I_0=0.05$, and $\lambda=1$, chosen to put the $A$ and $I$ contributions on comparable scales for the projectiles considered.

A two-point interpolation factor is then introduced,
\begin{equation}
\alpha_P=\frac{D_{P,78}}{D_{P,78}+D_{P,124}},
\label{eq:alphaP}
\end{equation}
and an effective objective-function value for projectile $P$ is obtained as
\begin{equation}
\mathcal{J}^{(P)}_{m}=(1-\alpha_P)\,\mathcal{J}^{(78)}_{m}+\alpha_P\,\mathcal{J}^{(124)}_{m}.
\label{eq:Jinterp}
\end{equation}
To convert the interpolated objective-function values into relative model weights for projectile $P$, we adopt an exponential mapping,
\begin{equation}
L^{(P)}_{m}=\exp\!\left(-\mathcal{J}^{(P)}_{m}/2\right),
\label{eq:Linterp}
\end{equation}
and the normalized model-averaging weight is
\begin{equation}
w^{(P)}_{m}=\frac{L^{(P)}_{m}}{\sum\limits_{n}L^{(P)}_{n}} .
\label{eq:w_norm}
\end{equation}
This mapping is used for the propagated systems to provide a monotonic reweighting that more strongly suppresses models with larger $\mathcal{J}^{(P)}_{m}$, whereas the calibration reactions use the inverse-objective weighting defined in Sec.~\ref{sec:BMA_78Kr_124Xe}.
Equations~\eqref{eq:dist_AI}--\eqref{eq:w_norm} are applied for $P=\nuc{92}{Mo}$ and $P=\nuc{144}{Sm}$. Numerical values of $D_{P,i}$ and $\alpha_P$ are provided in Appendix~\ref{app:weights_interp}.

The parameters $S_{\sigma}$, $\delta_{oe}$, and $\Delta R$ for $P=\nuc{92}{Mo}$ and $P=\nuc{144}{Sm}$ were obtained using the same two-point interpolation scheme as in Eq.~\eqref{eq:Jinterp}. For $P=\nuc{144}{Sm}$, the odd--even parameter $\delta_{oe}$ was not interpolated but set equal to the \nuc{124}{Xe} value, reflecting the assumption that the effective odd--even correction does not increase further for heavier projectiles; within the present calibration uncertainty, this choice is consistent with the default setting ($\delta_{oe}=12$ MeV).
\subsection{Mass-model parameter sets for propagated projectiles}
\label{sec:param_tables}

For practical reproducibility and direct use as \lisepp\ AA inputs, the mass-model--dependent
parameter sets adopted for the \nuc{92}{Mo} and \nuc{144}{Sm} projectiles are compiled in
Tables~\ref{tab:92mo} and \ref{tab:144sm}. For each mass model $m$, the excitation-energy
parameters $(k_1,k_2,k_{NZ})$ entering Eq.~\eqref{eq:Tmean} are obtained from the projectile
propagation procedure described in Sec.~\ref{sec:scale_e0}. The auxiliary parameters
$S_{\sigma}$, $\Delta R$, and $\delta_{oe}$ are assigned as described in Sec.~\ref{sec:weights_scale};
in particular, for $P=\nuc{144}{Sm}$ the odd--even parameter $\delta_{oe}$ is taken from the
corresponding \nuc{124}{Xe} fit for each mass model (no additional interpolation).
The weighted averages and dispersions in the last two rows are recalculated from these projectile-specific parameter sets using the corresponding weights $w_m^{(P)}$.

The column labeled $w_m^{(P)}$ lists the normalized BMA weights for each projectile, obtained from the
interpolated objective function $\mathcal{J}^{(P)}_m$ [Eq.~\eqref{eq:Jinterp}] via
$L^{(P)}_{m}=\exp\!\left(-\mathcal{J}^{(P)}_{m}/2\right)$ and
$w^{(P)}_{m}=L^{(P)}_{m}/\sum_n L^{(P)}_{n}$ [Eq.~\eqref{eq:w_norm}], such that $\sum_m w_m^{(P)}=1$.
Here BMA is used in the Bayesian-inspired sense of weighted averaging over AA realizations, with weights derived from the objective-function values through the mapping defined above.
These weights are used to construct the model-averaged (BMA) cross sections,
\begin{equation}
\sigma_{\mathrm{BMA}}^{(P)}(A,Z)=\sum_{m} w^{(P)}_{m}\,\sigma^{(P)}_{m}(A,Z),
\label{eq:sigma_bma}
\end{equation}
with the associated model-spread uncertainty taken as the weighted standard deviation,
\begin{equation}
\Delta\sigma^{(P)}_{m}(A,Z)\equiv
\sigma^{(P)}_{m}(A,Z)-\sigma_{\mathrm{BMA}}^{(P)}(A,Z),
\end{equation}
\begin{equation}
\delta\sigma_{\mathrm{BMA}}^{(P)}(A,Z)=
\left[\sum_{m} w^{(P)}_{m}\left(\Delta\sigma^{(P)}_{m}(A,Z)\right)^2\right]^{1/2}.
\label{eq:sigma_bma_std}
\end{equation}
The final two rows of Tables~\ref{tab:92mo} and \ref{tab:144sm} report the weighted averages of the adopted AA parameter sets and their weighted dispersions.

\begin{table}[htbp]
\centering
\setlength{\extrarowheight}{1.5pt}
\setlength{\tabcolsep}{2.0pt}

\caption{Mass-model--dependent AA parameter sets used for \nuc{92}{Mo} projectile-fragmentation calculations. The BMA weights $w_m$ are normalized such that $\sum_m w_m=1$.}
\label{tab:92mo}
\small
\begin{ruledtabular}
\begin{tabular}{lc|
>{\footnotesize}c|
>{\footnotesize}r
>{\footnotesize}r
>{\footnotesize}r|
>{\footnotesize}r
>{\footnotesize}r
>{\footnotesize}r}
\multicolumn{2}{c|}{Mass model} & $w_m^{(92)}$ & $k_1$ & $k_2$ & $k_{NZ}$   & $S_{\sigma}$ & $\delta_{oe}$ & $\Delta R$ \\
\hline
AME2020   & \cite{AME2020}    & 0.114 & 14.6 & -0.42 &  0.32 & 1.0 &  9.6 & 5.4 \\
FRDM2012  & \cite{PM-PRL12}   & 0.094 & 13.9 & -0.47 &  0.48 & 1.3 &  8.2 & 5.8 \\
HFB-22    & \cite{SG-PRC13}   & 0.061 & 16.5 & -0.24 &  0.09 & 1.3 & 10.0 & 5.9 \\
HFB-27    & \cite{SG-PRC13}   & 0.106 & 16.1 & -0.46 &  0.33 & 0.8 &  8.3 & 5.8 \\
KTUY      & \cite{KTUY-PTP05} & 0.141 & 14.5 & -0.47 &  0.40 & 1.2 & 10.1 & 5.9 \\
LISE-LDM3 & \cite{LISE08}     & 0.072 & 15.6 & -0.36 &  0.34 & 1.5 &  9.0 & 5.7 \\
NL3*      & \cite{rN3}        & 0.023 & 16.2 &  0.15 & -0.10 & 1.3 & 10.8 & 5.9 \\
SLy4      & \cite{SLy4}       & 0.061 & 17.7 & -0.18 &  0.36 & 1.3 & 11.5 & 5.7 \\
SV-min    & \cite{SVmin}      & 0.078 & 16.1 & -0.07 &  0.43 & 0.9 & 12.1 & 5.5 \\
TUYY      & \cite{TTYY}       & 0.080 & 14.6 & -0.4  &  0.28 & 1.0 &  8.9 & 5.7 \\
UNEDF1    & \cite{UNEDF1}     & 0.063 & 13.2 &  0.00 &  0.39 & 0.8 & 11.3 & 5.8 \\
WS4$_{\mathrm{RBF}}$ & \cite{WS4RBF} & 0.107 & 14.5 & -0.45 &  0.34 & 0.8 &  9.0 & 5.6 \\
\hline
\multicolumn{3}{r|}{\textit{Weighted average \hspace*{0.4cm}}}   & 15.0 & -0.38 & 0.36 & 1.0 &  9.6 & 5.7 \\
\multicolumn{3}{r|}{\textit{Uncertainty \hspace*{0.4cm}}}      &  0.1 &  0.14 & 0.09 & 0.2 &  1.1 & 0.2 \\
\end{tabular}
\end{ruledtabular}
\end{table}

\begin{table}[htbp]
\centering
\setlength{\extrarowheight}{1.5pt}
\setlength{\tabcolsep}{2.0pt}

\caption{Mass-model--dependent AA parameter sets used for \nuc{144}{Sm} projectile-fragmentation calculations. The BMA weights $w_m^{(144)}$ are normalized such that $\sum_m w_m^{(144)}=1$.}
\label{tab:144sm}
\small
\begin{ruledtabular}
\begin{tabular}{lc|
>{\footnotesize}c|
>{\footnotesize}r
>{\footnotesize}r
>{\footnotesize}r|
>{\footnotesize}r
>{\footnotesize}r
>{\footnotesize}r}
\multicolumn{2}{c|}{Mass model} & $w_m^{(144)}$ & $k_1$ & $k_2$ & $k_{NZ}$ & $S_{\sigma}$ & $\delta_{oe}$ & $\Delta R$ \\
\hline
AME2020   & \cite{AME2020}    & 0.094 & 27.1 & -0.53 &  0.56 & 0.6 & 12.0 & 4.2 \\
FRDM2012  & \cite{PM-PRL12}   & 0.066 & 24.1 & -0.34 &  0.26 & 0.7 &  9.4 & 4.3 \\
HFB-22    & \cite{SG-PRC13}   & 0.074 & 25.5 & -0.10 & -0.37 & 1.6 & 12.2 & 5.6 \\
HFB-27    & \cite{SG-PRC13}   & 0.147 & 23.8 & -0.14 & -0.20 & 0.8 & 11.7 & 4.7 \\
KTUY      & \cite{KTUY-PTP05} & 0.131 & 23.6 & -0.46 &  0.42 & 0.9 & 13.0 & 4.7 \\
LISE-LDM3 & \cite{LISE08}     & 0.081 & 22.4 & -0.52 & -0.11 & 1.6 & 11.2 & 5.0 \\
NL3*      & \cite{rN3}        & 0.007 & 21.0 & -0.18 &  0.42 & 1.7 & 12.5 & 5.7 \\
SLy4      & \cite{SLy4}       & 0.035 & 23.6 & -0.52 &  0.55 & 1.5 & 13.0 & 5.3 \\
SV-min    & \cite{SVmin}      & 0.071 & 28.0 & -0.36 &  0.53 & 0.8 & 13.0 & 4.3 \\
TUYY      & \cite{TTYY}       & 0.074 & 23.0 & -0.30 &  0.15 & 1.0 & 12.9 & 5.0 \\
UNEDF1    & \cite{UNEDF1}     & 0.100 & 22.0 & -0.40 &  0.33 & 1.3 & 13.0 & 5.2 \\
WS4$_{\mathrm{RBF}}$ & \cite{WS4RBF} & 0.119 & 21.3 & -0.02 & -0.09 & 0.9 & 12.9 & 4.8 \\
\hline
\multicolumn{3}{r|}{\textit{Weighted average \hspace*{0.4cm}}} & 23.7 & -0.29 & 0.12 & 1.0 & 12.3 & 4.8 \\
\multicolumn{3}{r|}{\textit{Uncertainty \hspace*{0.4cm}}}     &  1.8 &  0.18 & 0.31 & 0.3 &  0.9 & 0.4 \\
\end{tabular}
\end{ruledtabular}
\end{table}

\subsection{Model-averaged cross sections for \nuc{92}{Mo} and \nuc{144}{Sm}}
\label{sec:MoSm_fig}

\begin{figure*}[htbp]
\centering
\begin{subfigure}[t]{0.495\textwidth}
  \centering
  \includegraphics[width=\linewidth]{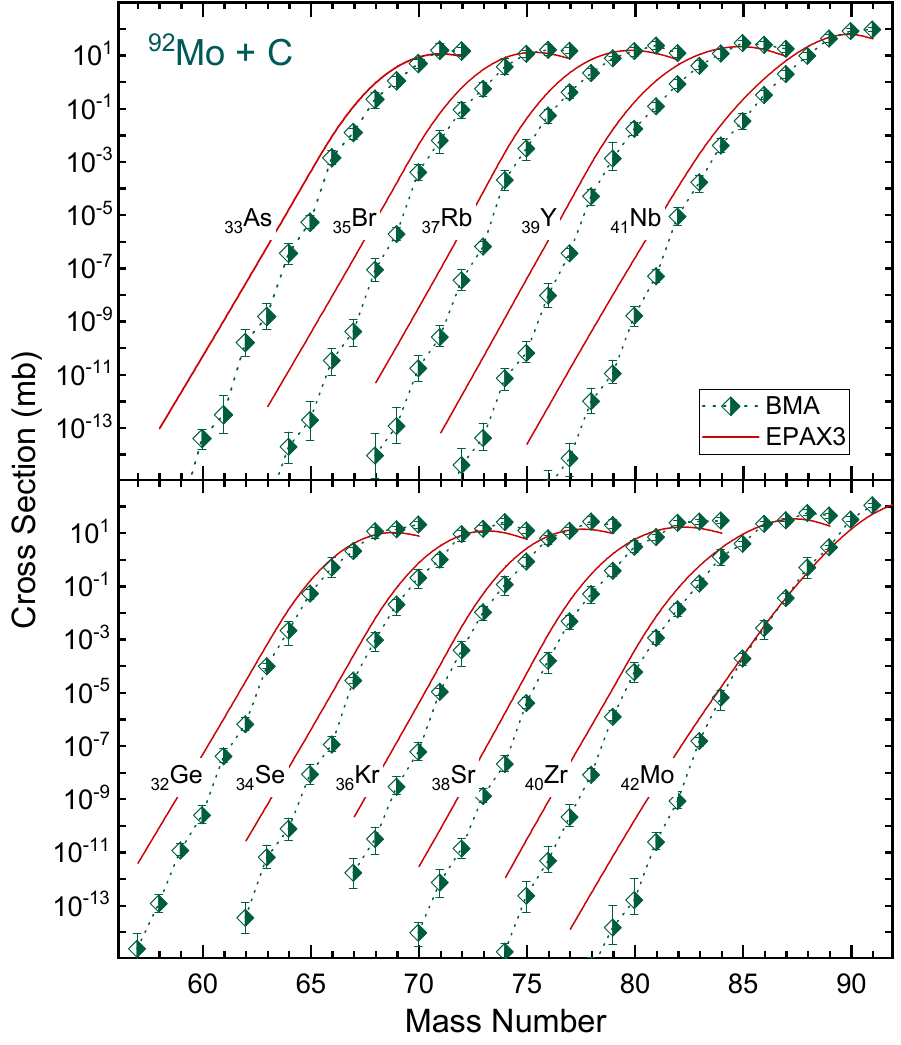}
\end{subfigure}
\hfill
\begin{subfigure}[t]{0.495\textwidth}
  \centering
  \includegraphics[width=\linewidth]{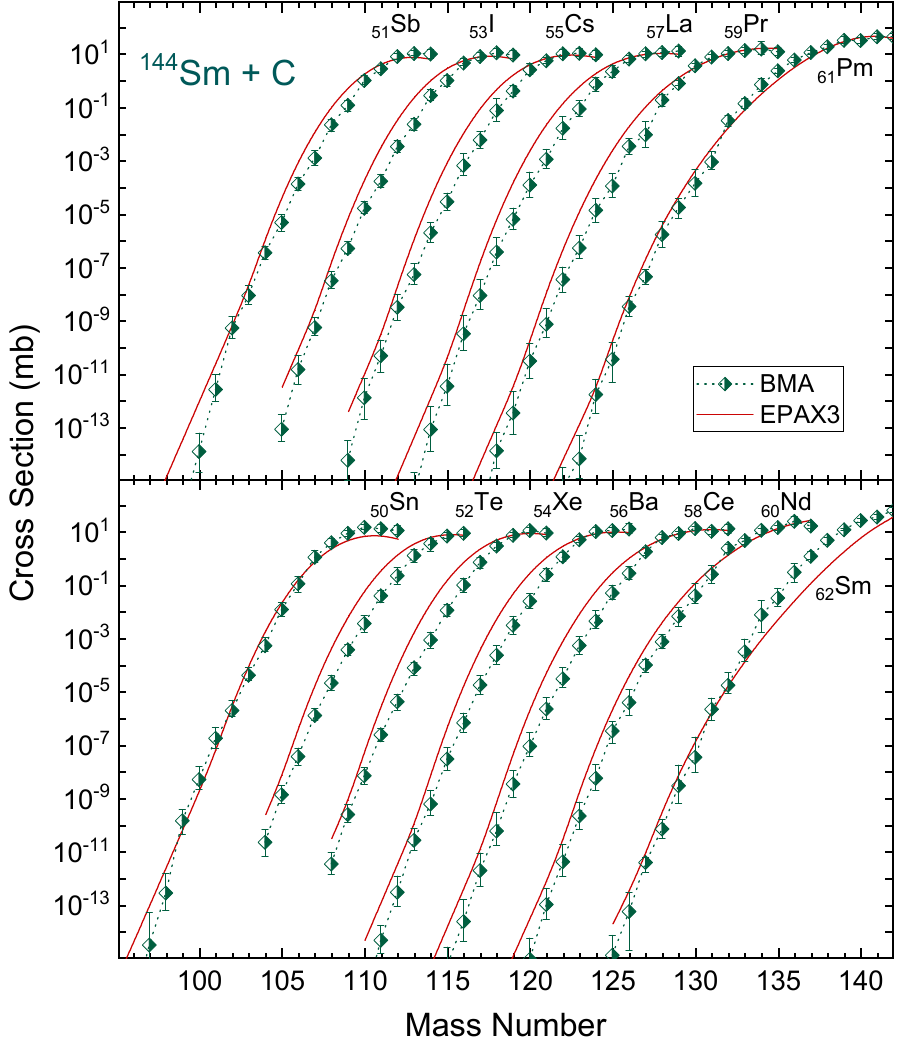}
\end{subfigure}
\caption{\label{fig:MoSm_BMA}
Model-averaged abrasion--ablation (BMA-inspired) production cross sections compared with the
EPAX3 parametrization. Isotopic distributions are shown as a function of mass number on a
logarithmic cross-section scale. The left panel corresponds to the \nuc{92}{Mo}+\nuc{12}{C} reaction for
fragments in the range $32 \le Z \le 42$, while the right panel corresponds to the
\nuc{144}{Sm}+\nuc{12}{C} reaction for fragments in the range $50 \le Z \le 62$. In both cases, the
BMA predictions are obtained by propagating the trends and weights constrained by the
\nuc{78}{Kr} and \nuc{124}{Xe} analyses.}
\end{figure*}

Using the mass-model parameter sets and normalized weights $w_m^{(P)}$ summarized in
Tables~\ref{tab:92mo} and \ref{tab:144sm}, AA calculations were performed for each mass model
to obtain isotope-production cross sections $\sigma_m^{(P)}(A,Z)$ for $P=\nuc{92}{Mo}$ and
$P=\nuc{144}{Sm}$. The corresponding model-averaged predictions,
$\sigma_{\mathrm{BMA}}^{(P)}(A,Z)$, were then constructed according to Eq.~\eqref{eq:sigma_bma},
with model-spread uncertainties from Eq.~\eqref{eq:sigma_bma_std}. Figure~\ref{fig:MoSm_BMA}
shows the resulting isotopic distributions and compares them with the EPAX3 parametrization,
illustrating the impact of the BMA-inspired averaging when trends constrained by the
\nuc{78}{Kr} and \nuc{124}{Xe} analyses are applied to additional projectiles. Although the
calibration data are for \nuc{9}{Be} targets, the \nuc{92}{Mo} and \nuc{144}{Sm} predictions are
evaluated for a carbon production target to match typical FRIB beam-production conditions.

In Fig.~\ref{fig:MoSm_BMA}, the propagated BMA predictions exhibit a markedly different behavior
relative to EPAX3 for the two projectiles. For \nuc{92}{Mo}+C, EPAX3 tends to
produce broader isotopic distributions with comparatively enhanced proton-rich tails, whereas the
BMA-inspired AA calculations predict a steeper suppression toward the most exotic (lowest-yield)
isotopes. In contrast, for \nuc{144}{Sm}+C, the BMA and
EPAX3 systematics are generally closer over most isotopic chains, with visible differences mainly
restricted to the extreme low-cross-section region.
This trend mirrors the behavior observed in the calibration reactions: EPAX3 shows larger
systematic deviations in the tails of the \nuc{78}{Kr}+\nuc{9}{Be} data, while the agreement is
typically better for \nuc{124}{Xe}+\nuc{9}{Be}.
Accordingly, when the excitation-energy trends and mass-model weights constrained by
\nuc{78}{Kr} and \nuc{124}{Xe} are propagated to additional projectiles, the resulting corrections
to EPAX3-like behavior are more pronounced for \nuc{92}{Mo} than for \nuc{144}{Sm}.


To provide a global cross-check of the propagated BMA-inspired AA predictions, Appendix~\ref{app:full_dists} presents a simultaneous comparison with the EPAX3 parametrization for all four projectiles considered in this work. Full isotopic distributions are shown for representative fragment chains with $Z=Z_{\rm proj}-8$, which avoids the truncation used in  Figs.~\ref{fig:BMAcompare} and \ref{fig:MoSm_BMA} to prevent overlap and makes the
neutron-rich side visible. Overall, this global comparison provides a compact consistency check of the systematics on both sides of the isotopic distributions.


\section{Discussion}
\label{sec:discussion}

\subsection{Beam selection for proton-rich \textit{Z}=50, 52, and 54 isotopes}
\label{subsec:beamSelection_Z505254}

\begin{figure}[t]
\centering
  \includegraphics[width=0.99\linewidth]{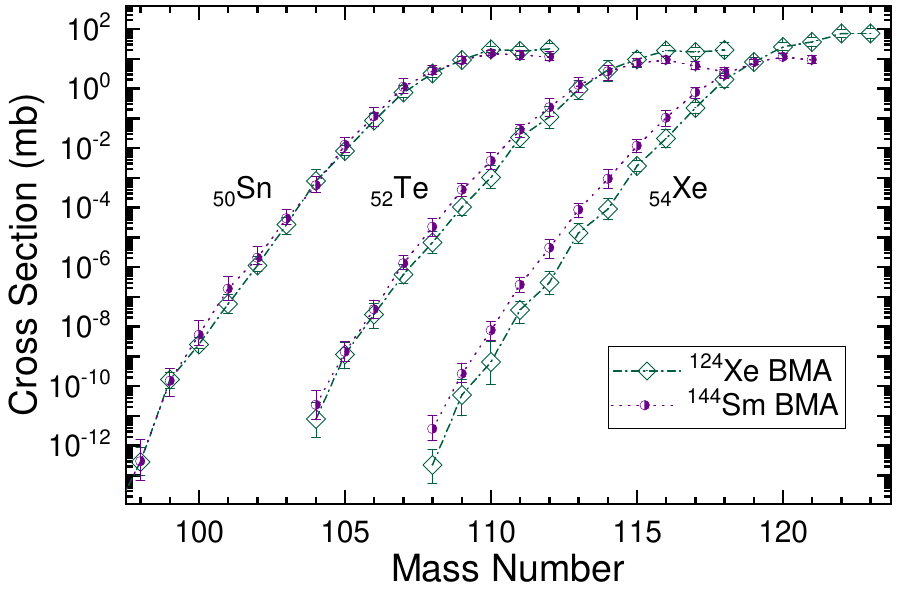}
  \caption{\label{fig:XeSm_Z505254}
BMA-inspired AA production cross sections for fragments with $Z=50$, 52, and 54, comparing projectile choices (\nuc{124}{Xe} and \nuc{144}{Sm}). The comparison illustrates how propagated BMA cross sections can guide projectile selection for targeted isotope production when combined with separator acceptance and transmission.}
\end{figure}

The production of proton-rich isotopes in the $Z=50$--54 region is strongly constrained by the choice of primary beam. 
At intermediate energies, the attainable yields depend not only on the absolute fragmentation cross sections, but also on how efficiently the separator can transmit the nuclei of interest while suppressing competing reaction products. 
Figure~\ref{fig:XeSm_Z505254} summarizes this interplay by comparing the BMA-inspired abrasion--ablation production cross sections for fragments with $Z=50$, 52, and 54 for the candidate projectiles considered in this work.

For the present discussion, the projectile choice is guided primarily by the expected \emph{production rate}. 
No additional ``purity'' criterion (often favoring fragments closer to the projectile) is applied; the comparison focuses instead on maximizing the yield of the nuclei of interest. 
Potential differences in multistep-reaction contributions---which may be slightly enhanced for more dissipative pathways and could therefore favor the \nuc{144}{Sm} projectile---are also not considered in this comparison.

Besides the production cross section, the separator transmission must be taken into account. 
In general, fragments produced far from the projectile exhibit broader effective distributions, reducing the fraction transmitted through a finite-acceptance separator. 
Two effects dominate: (i) the intrinsic longitudinal-momentum width increases with the amount of mass removed, following the usual systematics $\sigma_p \propto \sqrt{\Delta A}$; and (ii) for a thick production target the reaction may occur anywhere along the target thickness, so the fragments emerge with a finite energy (and thus $B\rho$) window determined by the reaction-depth distribution. 
The size of this ``energy rectangle'' is controlled by the difference in stopping power between projectile and fragment and therefore grows with $\Delta Z$, leading to a broader range of fragment energies at the target exit. 
Consequently, projectiles closer in $(A,Z)$ to the nucleus of interest generally provide a transmission advantage. 
This effect is non-negligible even for large-acceptance devices such as ARIS~\cite{ARIS2013,ARIS2023}; for example, at $\sim 200$~MeV/u the calculated transmission of \nuc{111}{Xe} is about a factor of 1.35 higher for a \nuc{124}{Xe} projectile than for \nuc{144}{Sm}.

To quantify the relative advantage of \nuc{144}{Sm} versus \nuc{124}{Xe} in the extreme low-cross-section region, projectile-to-projectile gain factors were evaluated using the  model-averaged (BMA) cross sections in Fig.~\ref{fig:XeSm_Z505254} and are summarized in Table~\ref{tab:gain_selected}. 
For a selected fragment $F(A,Z)$, the gain factor is defined as
\begin{equation}
R_{F,0} \equiv
\frac{\sigma_F(\nuc{144}{Sm}+\nuc{12}{C})}{\sigma_F(\nuc{124}{Xe}+\nuc{12}{C})}.
\end{equation}
The ratio uncertainty is constructed conservatively from the asymmetric BMA model-spread values
$\delta\sigma^{+}$ and $\delta\sigma^{-}$ by bracketing
\begin{equation}
R_F^{\rm hi}=\frac{\sigma_F^{\rm Sm}+\delta\sigma_{F,{\rm Sm}}^{+}}{\sigma_F^{\rm Xe}-\delta\sigma_{F,{\rm Xe}}^{-}},
\qquad
R_F^{\rm lo}=\frac{\sigma_F^{\rm Sm}-\delta\sigma_{F,{\rm Sm}}^{-}}{\sigma_F^{\rm Xe}+\delta\sigma_{F,{\rm Xe}}^{+}},
\end{equation}
and quoting $R_F=R_{F,0}\,^{+\,(R_F^{\rm hi}-R_{F,0})}_{-\,(R_{F,0}-R_F^{\rm lo})}$. \vspace{0.2 cm}

\begin{table}[tbp]
\centering
\setlength{\extrarowheight}{2.0pt}
\setlength{\tabcolsep}{2.5pt}
\caption{Projectile-to-projectile gain factors for selected proton-rich nuclei,
chosen to span the $\sim10^{-11}$--$\sim10^{-8}$\,mb region of the model-averaged (BMA) cross sections.
Asymmetric uncertainties are constructed from the BMA model-spread values (see text).}\label{tab:gain_selected}
\setlength{\tabcolsep}{6pt}
\renewcommand{\arraystretch}{1.15}
{\small
\begin{tabular}{lcc}
\hline
Element & Selected isotope (scale) & $R_F=\sigma_{\rm Sm}/\sigma_{\rm Xe}$ \\
\hline
Sn & $^{100}$Sn ($\sim10^{-8}$\,mb)  & $2.12^{+7.1}_{-1.6}$ \\
   & $^{99}$Sn  ($\sim10^{-10}$\,mb) & $0.902^{+3.9}_{-0.75}$ \\
\hline
Te & $^{106}$Te ($\sim10^{-8}$\,mb)  & $1.52^{+7.8}_{-1.2}$ \\
   & $^{104}$Te ($\sim10^{-11}$\,mb) & $2.97^{+35}_{-2.7}$ \\
\hline
Xe & $^{111}$Xe ($\sim10^{-8}$\,mb)  & $7.12^{+29}_{-5.2}$ \\
   & $^{109}$Xe ($\sim10^{-11}$\,mb) & $5.12^{+56}_{-4.4}$ \\
\hline
\end{tabular}
}
\end{table}

In summary, combining the relative production cross sections and transmission factors---and, potentially, qualitative purity considerations---leads to a hierarchy among the candidate beams, although the ranking becomes less decisive once the BMA model-spread uncertainties and separator transmission are taken into account. The \nuc{144}{Sm} projectile is favored for the production of the most proton-rich Xe isotopes, where the gain in cross section can compensate for somewhat reduced transmission. Its advantage becomes less pronounced for Te isotopes, where the balance between yield and transmission is more comparable.
For Sn isotopes, the \nuc{124}{Xe} beam is favored (within uncertainties), reflecting its closer proximity in $(A,Z)$ and typically more favorable transmission conditions.

\subsection{Implications for new-isotope searches}
\label{sec:disc_newiso}

Most recent discoveries of new proton-rich isotopes in the region $30 \le Z \le 70$ have been achieved using a limited set of primary beams, most notably \nuc{78}{Kr}, \nuc{124}{Xe}, and \nuc{238}{U}. While \nuc{78}{Kr} and \nuc{124}{Xe} provide excellent access to broad areas on the proton-rich side, their relatively low $Z$ leaves ``blind spots'' at higher $Z$, where production becomes increasingly unfavorable; \nuc{238}{U} is largely exempt from this limitation due to its high charge and mass. The use of intermediate-$Z$ projectiles such as \nuc{92}{Mo} and \nuc{144}{Sm} can help bridge these gaps and improve coverage of proton-rich nuclei beyond the reach of \nuc{78}{Kr} and \nuc{124}{Xe}. 

In the present analysis, attention is restricted to unobserved proton-rich isotopes in regions that are not efficiently populated by \nuc{78}{Kr} and \nuc{124}{Xe}. Despite their non-observation to date, a first requirement is to assess whether the candidate nuclei are experimentally accessible through proton radioactivity, i.e., whether their expected half-lives fall within a range compatible with observation and identification in the separator and detector system.

Proton-radioactivity half-lives were estimated using the single-particle $R$-matrix formalism of Brown and Barker~\cite{BrownBarker2003_diproton45Fe}, as implemented in the \soft{proton\_f} utility and incorporated into \liseppsh. In this approach, the partial decay width is computed from the $R$-matrix penetrability (including the associated shift function) and a reduced width proportional to the spectroscopic factor. Partial widths were evaluated for both one-proton and two-proton emission channels when energetically allowed, using the corresponding $Q_{p}$ and $Q_{2p}$ values.
For a conservative estimate of observability, the effective half-life was taken as
$T_{1/2}=\min(T_{1/2}^{(p)},\,T_{1/2}^{(2p)}).$
For the decay $Q$ values entering the $T_{1/2}$ calculations, the KTUY nuclear-mass model~\cite{KTUY-PTP05} was adopted for the \nuc{92}{Mo} case, since it carries the highest statistical weight among the considered mass tables in the BMA analysis (see Table~\ref{tab:92mo}). The HFB-27 model~\cite{SG-PRC13} was adopted accordingly for the \nuc{144}{Sm} case (see Table~\ref{tab:144sm}).

For the rate estimates, primary-beam intensities corresponding to $I=1.0$ and $0.5~p\mu$A were assumed for the \nuc{92}{Mo} and \nuc{144}{Sm} beams, respectively. 
Representative fragment transmissions in the relevant production regions were taken as 60\% for \nuc{92}{Mo} and 15\% for \nuc{144}{Sm}. 
A carbon production target of 3.5~mm thickness was assumed, with a density of $\rho=1.89~\mathrm{g/cm^3}$.
Tables~\ref{tab:rates_t12_92Mo} and \ref{tab:rates_t12_144Sm} summarize the estimated production rates and proton-radioactivity half-lives for selected $Z$ candidates populated with the \nuc{92}{Mo} and \nuc{144}{Sm} primary beams. The rates were derived using the BMA-inspired abrasion–ablation production cross sections discussed above, combined with the assumed primary-beam intensity and separator transmission. The listed half-lives were calculated within the single-particle R-matrix framework described previously, using the spherical valence-proton configurations indicated in the table.



\begin{table}[tbp]
\centering
\setlength{\extrarowheight}{1.5pt}
\renewcommand{\arraystretch}{1.15}
\setlength{\tabcolsep}{6pt}
\caption{Estimated production rates (per day) and proton-radioactivity half-lives for selected candidates populated with the \nuc{92}{Mo} beam.
Half-life estimates were calculated using the KTUY nuclear-mass table~\cite{KTUY-PTP05}. The assumed spherical valence-proton orbital ($\pi n\ell_j$) and the corresponding orbital angular momentum $\ell$
are indicated. For \nuc{81}{Nb}, two alternative $p$-emission angular-momentum assumptions ($\ell=1$ and $\ell=4$)
are listed, together with the experimental upper limit.}
\label{tab:rates_t12_92Mo}
\begin{tabular}{ccccc}
\hline
Isotope & Rate (day$^{-1}$)  & $T_{1/2}$ (s) & $\pi n\ell_j$ & $\ell$ \\
\hline
\nuc{70}{Sr} & 0.11  & $4.1\times10^{-9}$ & \multirow{3}{*}{$\pi\,1f_{5/2}$} & \multirow{3}{*}{3} \\
\nuc{71}{Sr} & 8.76  & $>1$               &  &  \\
\nuc{72}{Sr} & 164.9 & $>1$               &  &  \\
\hline
\nuc{75}{Y}$^{\mathrm{a}}$ & 800   & $1.2\times10^{-12}$ & $\pi\,1g_{9/2}$ & 4 \\
\hline
\nuc{74}{Zr} & 0.02  & $2.3\times10^{-8}$ & \multirow{3}{*}{$\pi\,2p_{1/2}$} & \multirow{3}{*}{1} \\
\nuc{75}{Zr} & 2.74  & $2.3\times10^{-1}$ &  &  \\
\nuc{76}{Zr} & 57.5  & $>1$               &  &  \\
\hline
\multirow{3}{*}{\nuc{81}{Nb}} &
\multirow{3}{*}{$6\times10^{5}$} &
$3.1\times10^{-7}$ & $\pi\,2p_{1/2}$ & 1 \\
& & $4.0\times10^{-4}$ & $\pi\,1g_{9/2}$ & 4 \\
& & \multicolumn{3}{l}{$<4.0\times10^{-8}$ (exp. limit) \cite{72Rb_Suzuki}} \\
\hline
\nuc{78}{Mo} & 0.004 & $8.7\times10^{-8}$ & \multirow{3}{*}{$\pi\,1g_{9/2}$} & \multirow{3}{*}{4} \\
\nuc{79}{Mo} & 0.176 & $7.5\times10^{-3}$ &  &  \\
\nuc{80}{Mo} & 1.92  & $>1$               &  &  \\
\hline
\end{tabular}
\vspace{2pt}
\par\small
\noindent\justify $^{\mathrm{a}}$~The isotope $^{75}$Y does not yet have a directly measured half-life; the evaluated NUBASE2020 compilation lists an estimated value of $T_{1/2}=100^{\#}\ \mu$s with tentative $J^\pi = 5/2^{+\#}$ and uncertain decay branches~\cite{NUBASE2020}. A dedicated search
at RIKEN (BigRIPS) reported no identified \nuc{75}{Y} events, and therefore could not extract an
experimental $T_{1/2}$ \cite{RIBF97_75Ysearch}.
\end{table}

\begin{table}[tbp]
\centering
\setlength{\extrarowheight}{1.5pt}
\renewcommand{\arraystretch}{1.15}
\setlength{\tabcolsep}{6pt}
\caption{Estimated production rates (per day) and proton-radioactivity half-lives for selected candidates populated with the \nuc{144}{Sm} beam. Half-life estimates were calculated using HFB-27 mass table~\cite{SG-PRC13}. The adopted spherical valence-proton orbital ($\pi n\ell_j$) and $\ell$ follow the $50<Z<82$ ordering $\pi g_{7/2}$ ($\ell=4$) then $\pi d_{5/2}$ ($\ell=2$). For \nuc{125}{Pm}, an alternative $\ell=4$ estimate is also listed in italics.}
\label{tab:rates_t12_144Sm}
\begin{tabular}{ccccc}
\hline
Isotope & Rate (day$^{-1}$) & $T_{1/2}$ (s) & $\pi n\ell_j$ & $\ell$ \\
\hline
\nuc{111}{Cs} &  61.9 & $2.3\times10^{-12}$ & {$\pi\,g_{7/2}$} & 4 \\
\hline
\nuc{112}{Ba} & 0.39 & $>1$ & \multirow{2}{*}{$\pi\,g_{7/2}$} & \multirow{2}{*}{4} \\
\nuc{113}{Ba} & 34.9 & $>1$ &  &  \\
\hline
\nuc{115}{La} & 4.50  & $4.5\times10^{-13}$ & $\pi\,g_{7/2}$ & 4 \\

\hline
\nuc{116}{Ce} & 0.31 & $>1$ & \multirow{3}{*}{$\pi\,g_{7/2}$} & \multirow{3}{*}{4} \\
\nuc{117}{Ce} & 2.51 & $>1$ &  &  \\
\nuc{118}{Ce} & 77.8 & $>1$ &  &  \\
\hline
\nuc{119}{Pr} & 0.45   & $8.8\times10^{-14}$ & \multirow{2}{*}{$\pi\,d_{5/2}$} & \multirow{2}{*}{2} \\
\nuc{120}{Pr} & 39.3   & $1.3\times10^{-10}$ &  &  \\
\hline
\nuc{121}{Nd} & 0.13 & $>1$ & \multirow{2}{*}{$\pi\,d_{5/2}$} & \multirow{2}{*}{2} \\
\nuc{122}{Nd} & 5.42 & $>1$ &  &  \\
\hline
\nuc{124}{Pm} & 2.18   & $2.3\times10^{-12}$ & $\pi\,d_{5/2}$ & 2 \\
\nuc{125}{Pm}$^{\mathrm{b}}$ & 45.4 & $3.7\times10^{-9}$ & $\pi\,d_{5/2}$ & 2 \\
\textit{(alter.)} &  & $\mathit{2.7\times10^{-7}}$ & $\mathit{\pi\,g_{7/2}}$ & \textit{4} \\\hline
\nuc{126}{Sm} & 0.07 & $>1$ & \multirow{2}{*}{$\pi\,d_{5/2}$} & \multirow{2}{*}{2} \\
\nuc{127}{Sm} & 5.03 & $>1$ &  &  \\
\hline
\end{tabular}
\vspace{2pt}
\par\small
\noindent\justify $^{\mathrm{b}}$~A single event attributed to \nuc{125}{Pm} was recently reported by the BigRIPS collaboration~\cite{Suzuki25_Z60}. The expected proton-decay half-life of \nuc{125}{Pm} is strongly configuration dependent; in Ref.~\cite{125Pm_T12} it was evaluated within a nonadiabatic quasiparticle model as a function of deformation, decaying-state spin, and proton energy, with representative curves for emission from spherical single-particle configurations (e.g., $\pi d_{5/2}$, $\pi g_{7/2}$, $\pi h_{11/2}$). The study concludes that proton emission from \nuc{125}{Pm} could be measurable for $T_{1/2}\gtrsim 0.2~\mu$s.
\end{table}

Comparing the estimated daily production rates with the expected $T_{1/2}$ values provides a first assessment of experimental feasibility, highlighting nuclei for which both sufficient yield and measurable proton-decay lifetimes are anticipated. Overall, the rate and half-life estimates in Tables~\ref{tab:rates_t12_92Mo} and \ref{tab:rates_t12_144Sm} indicate that intermediate-$Z$ projectiles can access a nontrivial set of previously unobserved proton-rich nuclei in the ``blind-spot'' regions of \nuc{78}{Kr} and \nuc{124}{Xe}, dominated by even-$Z$ elements. Using a practical observability criterion of $\gtrsim 1$ ion/day under the assumed beam intensities and transmissions, five even-$Z$ candidates are reachable with the \nuc{92}{Mo} beam (\nuc{71,72}{Sr}, \nuc{75,76}{Zr},  \nuc{80}{Mo}) and five even-$Z$ candidates with the \nuc{144}{Sm} beam (\nuc{113}{Ba}, \nuc{117,118}{Ce}, \nuc{122}{Nd}, \nuc{127}{Sm}).

\section{Summary}

A Bayesian-inspired model-averaging (BMA) framework for abrasion--ablation calculations is developed to reduce systematic model dependence in fragmentation cross-section predictions. The approach combines AA calculations performed with 12 nuclear mass tables, assigning empirical weights constrained by measured isotopic distributions for the calibration reactions \nuc{78}{Kr}+\nuc{9}{Be} and \nuc{124}{Xe}+\nuc{9}{Be}. The resulting model-averaged cross sections are then propagated to additional projectiles relevant for FRIB beam production, with the propagated objective-function values converted to relative model weights through an exponential mapping and the corresponding model-spread uncertainties retained as asymmetric bands.

When applied to \nuc{92}{Mo}+\nuc{12}{C} and \nuc{144}{Sm}+\nuc{12}{C}, the propagated BMA-inspired AA predictions show projectile-dependent differences relative to EPAX3: for \nuc{92}{Mo}+C the most proton-rich tails are suppressed more steeply, whereas for \nuc{144}{Sm}+C the two systematics are generally closer and diverge mainly in the extreme low-cross-section region. In the $Z=50$--54 region, the projectile choice is shown to depend not only on production cross sections but also on separator transmission. Quantitatively, the most proton-rich Xe region favors \nuc{144}{Sm} through substantial gain factors, while the balance shifts toward \nuc{124}{Xe} for Sn isotopes.

Implications for new-isotope searches in ``blind-spot'' regions beyond the reach of \nuc{78}{Kr} and \nuc{124}{Xe} are explored by combining BMA-inspired production rates with proton-(and two-proton-)decay half-life estimates calculated using the Brown--Barker single-particle $R$-matrix formalism.
Under representative FRIB conditions (beam intensities and typical transmission assumptions), the \nuc{92}{Mo} case identifies five favorable even-$Z$ candidates with rates $\gtrsim 1$ day$^{-1}$ (\nuc{71,72}{Sr}, \nuc{75,76}{Zr}, \nuc{80}{Mo}), while the \nuc{144}{Sm} case yields five candidates with comparatively strong daily rates (\nuc{113}{Ba}, \nuc{117,118}{Ce}, \nuc{122}{Nd}, \nuc{127}{Sm}). Overall, the results indicate that intermediate-$Z$ projectiles can extend discovery reach and provide an uncertainty-aware basis for projectile selection and yield estimates for future FRIB new-isotope campaigns.

\begin{acknowledgments}
The author thanks Shane Watters and Holly Matthews for carefully reading the manuscript and for helpful comments. The author acknowledges the use of ChatGPT (OpenAI) for language editing and improvement of the manuscript text. All scientific content, analysis, and conclusions are the sole responsibility of the author.

This material is based upon work supported by the U.S. Department of Energy, Office of Science, Office of Nuclear Physics, and used resources of the Facility for Rare Isotope Beams (FRIB) Operations, a DOE Office of Science User Facility, under Award No.~DE-SC0023633. This work was also supported by the U.S. National Science Foundation under Grant No.~PHY-23-10078.
\end{acknowledgments}

\section{Data availability}
The model-averaged (BMA) cross-section tables generated in this work are available in a public Zenodo repository, Ref.~\cite{Tarasov_BMA_dataset}.


\appendix

\section{AA settings: parameter definitions and defaults}
\label{app:AA_settings}

Here $\Delta R$ is a phenomenological surface-separation (barrier-radius shift) parameter (in fm) that accounts for the fact that the effective Coulomb barrier is evaluated at a distance slightly different from the simple touching radius,
\begin{equation}
R_{\mathrm{touch}}=r_0\!\left(A_1^{1/3}+A_2^{1/3}\right).
\end{equation}
The global scaling factor $S_{\sigma}$ accounts for residual overall normalization differences between calculated and measured cross sections that are not captured by the shape parameters. Allowing $S_{\sigma}$ to vary prevents such effects from being absorbed into the excitation-energy parameters. All other AA parameters were kept at their \lisepp\ default values to avoid over-parameterization and to preserve established benchmarking.

The limiting-temperature parameters were also kept at their \lisepp\ default values, consistent with previous AA benchmarks in which the best-fit values remain close to the defaults and show only weak sensitivity within the present data constraints.

\section{Implementation of the logarithmic deviation term}
\label{app:softpow10}

The AA minimization includes a logarithmic deviation contribution based on the residual
\begin{equation}
d_i = \log_{10}\!\left(\frac{\sigma_{\mathrm{calc},i}}{\sigma_{\mathrm{exp},i}}\right).
\end{equation}
Instead of using the expectation value $\langle |d| \rangle$ (the mean absolute logarithmic deviation) directly, the implementation maps $|d_i|$ to a positive penalty through a
``soft'' version of $10^{|d|}$ in order to reduce the leverage of extreme outliers while preserving a
multiplicative interpretation for typical deviations, consistent with the general philosophy of robust loss
functions and bounded-influence estimators~\cite{Huber1981,Hampel2005}.

Specifically, the contribution accumulated for each point is
\begin{equation}
\ell_i = w_i\, S\!\left(|d_i|\right),
\qquad
w_i=\frac{\sigma_{\mathrm{exp},i}}{w_{1,i}},
\end{equation}
and the reported logarithmic metric is the weighted average
\begin{equation}
\mathcal{L}_{\log} = \frac{\sum_i \ell_i}{\sum_i w_i}
= \frac{\sum_i w_i\,S(|d_i|)}{\sum_i w_i}.
\label{eq:Llog_softApp}
\end{equation}
Here $w_{1,i}$ is the effective cross-section uncertainty used for weighting (experimental uncertainty when available, otherwise a
fractional uncertainty $f_{\rm err}\sigma_{\mathrm{exp},i}$, with a minimum floor to avoid divergent weights).
See Eq.~\eqref{eq:w1_def} for the explicit definition of $w_{1,i}$, including the floor $\sigma_{\min}$.

The mapping function $S(|d|)$ is defined as
\begin{equation}
S(|d|) = 10^{z(|d|)} - 1,
\end{equation}
with a piecewise definition in $\log_{10}$ space,
\begin{equation}
z(|d|)=
\begin{cases}
\gamma |d|, & \gamma |d|\le z_{\rm cap},\\[4pt]
z_{\rm cap}+\tau \ln\!\left(1+\gamma |d|-z_{\rm cap}\right), & \gamma |d|> z_{\rm cap},
\end{cases}
\label{eq:z_softcap}
\end{equation}
where $\gamma=0.8$, $z_{\rm cap}=10$, and $\tau=0.06$ in the present implementation; these parameters were set empirically to provide soft capping of extreme outliers while leaving typical deviations essentially unchanged.
For moderate deviations, $S(|d|)\approx 10^{0.8|d|}-1$ (close to a multiplicative factor penalty), whereas for very large deviations
the growth is strongly reduced by the logarithmic tail in Eq.~\eqref{eq:z_softcap}. This soft-capping prevents a small number of
poorly reproduced points from dominating the minimization while retaining sensitivity to systematic multiplicative trends.


\section{Two-point calibration of the size--isospin scaling law}
\label{app:twopoint_scale}

The parameters $(K,C)$ in the separable size--isospin scaling law of Eq.~\eqref{eq:eps_scaling}
are obtained analytically from two calibration projectiles (indexed 1 and 2). Using
Eq.~\eqref{eq:eps_scaling},
\begin{equation}
\begin{aligned}
y_1 &= K\, f(A_1)\,\bigl(1+C I_1\bigr),\\
y_2 &= K\, f(A_2)\,\bigl(1+C I_2\bigr).
\end{aligned}
\label{eq:calib_eqs}
\end{equation}
where $y_i\equiv \varepsilon(A_i,Z_i)$, $I_i=(N_i-Z_i)/A_i$, and $f(A)=A^{1/3}$. Defining
$a=f(A_1)$ and $b=f(A_2)$, form the ratio
\begin{equation}
r=\frac{y_2}{y_1}\,\frac{a}{b}.
\label{eq:rdef_app}
\end{equation}
From Eq.~\eqref{eq:calib_eqs},
\begin{equation}
r=\frac{1+C I_2}{1+C I_1}.
\label{eq:r_relation_app}
\end{equation}
Solving Eq.~\eqref{eq:r_relation_app} for $C$ yields
\begin{equation}
C=\frac{1-r}{r\,I_1-I_2},
\label{eq:Csolve_app}
\end{equation}
and substituting back into Eq.~\eqref{eq:calib_eqs} gives
\begin{equation}
K=\frac{y_1}{a\,(1+C I_1)}.
\label{eq:Ksolve_app}
\end{equation}
With $(K,C)$ determined, the excitation-energy scale for a third projectile $(A_3,Z_3)$ follows
directly from Eq.~\eqref{eq:eps_scaling} evaluated at $(A_3,Z_3)$.

\section{Interpolation distances and $\alpha$ factors used for weight propagation}
\label{app:weights_interp}

Table~\ref{tab:alpha_dist} lists the scaled distances $D_{P,i}$ in the $(A,I)$ plane and the
corresponding interpolation factors $\alpha_P$ used to propagate BMA weights from the
calibrators \nuc{78}{Kr} ($i=78$) and \nuc{124}{Xe} ($i=124$) to the additional projectiles
$P=\nuc{92}{Mo}$ and $\nuc{144}{Sm}$ (Sec.~\ref{sec:weights_scale}).

\begin{table}[!h]  
\centering
\caption{Scaled distances $D_{P,i}$ and interpolation factors $\alpha_P$
(Eqs.~\ref{eq:dist_AI} and \ref{eq:alphaP}) used to propagate BMA weights from \nuc{78}{Kr}
and \nuc{124}{Xe} to additional projectiles.}
\label{tab:alpha_dist}
\setlength{\tabcolsep}{6pt}
\renewcommand{\arraystretch}{1.15}
{\small
\begin{tabular}{lcccccc}
\hline
Projectile $P$ &  $I_P$ & $D_{P,78}$ & $D_{P,124}$ & $\alpha_P$ \\
\hline
\nuc{92}{Mo}  &  0.0870 & 0.403 & 1.161 & 0.258 \\
\nuc{144}{Sm} &  0.1389 & 2.064 & 0.537 & 0.793 \\
\hline
\end{tabular}
}
\end{table}


\section{Full isotopic distributions for $Z=Z_{\rm proj}-8$}
\label{app:full_dists}

\begin{figure*}[t]
\centering
\begin{minipage}[c]{0.75\textwidth}
  \centering
  \includegraphics[width=\linewidth]{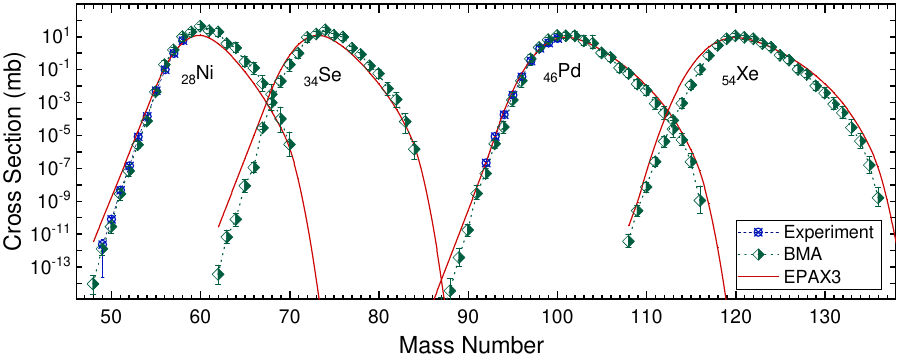}
\end{minipage}\hfill
\begin{minipage}[c]{0.23\textwidth}
\caption{\label{fig:Zminus8_full}
Global comparison of propagated BMA-inspired AA cross sections with the EPAX3 parametrization for all four
projectiles considered in this work. Full isotopic distributions are shown on a logarithmic scale as a function
of mass number for representative fragment chains with $Z=Z_{\rm proj}-8$, corresponding to
\nuc{78}{Kr}$\rightarrow$Ni, \nuc{92}{Mo}$\rightarrow$Se, \nuc{124}{Xe}$\rightarrow$Pd, and
\nuc{144}{Sm}$\rightarrow$Xe.
}
\end{minipage}
\end{figure*}

Figure~\ref{fig:Zminus8_full} provides a global comparison of the propagated BMA-inspired AA predictions with
the EPAX3 parametrization for all four projectiles analyzed in this work. Full isotopic distributions are shown
for representative fragment chains with $Z=Z_{\rm proj}-8$, allowing both the neutron-rich and proton-rich sides
to be inspected simultaneously and providing a compact consistency check of systematic trends across projectiles.

\bibliographystyle{apsrev4-2}

\end{document}